\documentclass[reprint,aps,prl,amsmath,amssymb,nofootinbib,superscriptaddress]{revtex4-1}

\usepackage[T1]{fontenc}
\usepackage[utf8]{inputenc}
\usepackage{graphicx,color,rotating,pifont}
\usepackage{mathtools}
\usepackage{siunitx}
\usepackage{bm}
\usepackage{ae}
\usepackage{float}
\usepackage{siunitx}
\usepackage{dcolumn}
\usepackage{txfonts}
\usepackage{tensor}
\usepackage{braket}
\usepackage{booktabs}
\usepackage{xspace}
\usepackage{mathrsfs}
\usepackage{amssymb} 
\usepackage{amsmath}
\usepackage{esvect}
\usepackage{diagbox} 

\usepackage{tabularx}  
\usepackage{soul}
\setstcolor{red}
\setul{}{.2ex} 

\usepackage[normalem]{ulem}

\usepackage{tikz}
\usetikzlibrary{positioning,calc,shapes,decorations.markings,decorations.pathmorphing}
\tikzset{asg/.cd,
  omega-vertex/.style={circle,solid,draw=black,fill=white,minimum size=5pt, inner sep=0pt},
  dbd-vertex/.style={coordinate},
  pline/.style={thick, postaction={decorate}, decoration={markings, mark=at position .5 with {\arrow[xshift=2pt]{stealth}}}},
  hline/.style={thick, postaction={decorate}, decoration={markings, mark=at position .5 with {\arrowreversed[xshift=-2pt]{stealth}}}},
  shift arrow/.style={/pgf/decoration/transform={xshift=#1}},
  shift arrow/.default=-2pt,
  dbd-2b/.style={decorate, decoration=snake},
  omega-2b/.style={densely dashed},
  neutron/.style={draw=blue},
  proton/.style={draw=red},
}

\usepackage[colorlinks,linkcolor=blue,anchorcolor=blue,citecolor=blue]{hyperref} 
\hypersetup{
   colorlinks=true, 
   linkcolor=blue,  
   citecolor=blue,  
   filecolor=blue,  
   urlcolor=blue    
}

\DeclareMathSymbol{\NS}{\mathord}{AMSb}{"4E}

\DeclareSIUnit{\fm}{\femto\meter}

\newcommand{\beq}{\begin{equation}}
\newcommand{\eeq}{\end{equation}}
\newcommand{\beqn}{\begin{eqnarray}}
\newcommand{\eeqn}{\end{eqnarray}}
\newcommand{\bsub}{\begin{subequations}}
\newcommand{\esub}{\end{subequations}}
\newcommand{\bpm}{\begin{pmatrix}}
\newcommand{\epm}{\end{pmatrix}}

\newcommand\identity{1\kern-0.25em\text{l}}

\begin{document}

\title{Subspace-projected multireference covariant density functional theory }

 \author{X. Zhang}   
 \affiliation{School of Physics and Astronomy, Sun Yat-sen University, Zhuhai 519082, P.R. China}     
 \affiliation{Department of Physics, Kyoto University,
Kyoto 606-8502, Japan}

   \author{C. C. Wang}   
  \affiliation{School of Physics and Astronomy, Sun Yat-sen University, Zhuhai 519082, P.R. China}

   \author{C. R. Ding}   
  \affiliation{School of Physics and Astronomy, Sun Yat-sen University, Zhuhai 519082, P.R. China}

  \author{J. M. Yao}   
  \email{Corresponding author: yaojm8@sysu.edu.cn}
  \affiliation{School of Physics and Astronomy, Sun Yat-sen University, Zhuhai 519082, P.R. China}    
  \affiliation{Guangdong Provincial Key Laboratory of Quantum Metrology and Sensing, Sun Yat-Sen University, Zhuhai 519082, China }

\date{\today}

\begin{abstract} 
Multireference density functional theory (MR-DFT) has been a pivotal method for studying nuclear low-lying states and neutrinoless double-beta ($0\nu\beta\beta$) decay. However, quantifying their theoretical uncertainties has been a significant challenge due to the computational demands. This study introduces a subspace-projected covariant density functional theory (SP-CDFT), which efficiently emulates MR-CDFT calculations for nuclear low-lying states. This approach leverages the eigenvector continuation method combined with the quantum-number projected generator coordinate method, based on a relativistic energy density functional (EDF). We apply SP-CDFT to investigate the correlations among the physical quantities of nuclear matter, nuclear low-lying spectroscopy, and the nuclear matrix elements (NMEs) of $0\nu\beta\beta$ decay in the two heaviest candidate nuclei. Our findings reveal generally strong correlations between the NMEs of $0\nu\beta\beta$ decay and the excitation energy of the $2_1^+$ state, as well as the $E2$ transition strength, although these correlations vary significantly among nuclei. This work also paves the way for refining nuclear EDF parameters using spectroscopic data.
\end{abstract}

 
\maketitle

\paragraph{Introduction.} 
Understanding nuclear spectroscopy is essential not only for probing the intricacies of strongly correlated quantum many-body systems~\cite{Ring:1980} but also for constraining fundamental symmetries and interactions at the nuclear-energy scale~\cite{Engel:2013PPNP,Yang:2023PPNP}. Theoretical calculations of nuclear matrix elements (NMEs) are pivotal for interpreting and designing experiments aimed at exploring new physics at the high-intensity frontier~\cite{Chupp:2019RMP,Snowmass_EDM:2022,Snowmass_NLDBD:2022,Agostini:2023}.  Among the various nuclear models, nuclear density functional theory (DFT) is uniquely versatile, successfully applied to study nuclear properties across a wide range of mass regions~\cite{Bender:2003RMP,Vretenar:2005PR,Meng:2005PPNP,Niksic:2011PPNP,Sheikh:2021qv}. This includes not only ground-state bulk properties~\cite{Goriely:2009PRL,Erler:2012,Afanasjev:2013,DRHBcMassTable:2024} but also nuclear low-lying spectroscopy~\cite{Bender:2004Global,Rodriguez:2014Global}, nuclear Schiff moments~\cite{Engel:2003PRC,Dobaczewski:2018PRL}, and $0\nu\beta\beta$ decay~\cite{Rodriguez:2010PRL,Mustonen:2013,Song:2014,Yao:2015,Ding:2023,Wang:2024PRC,Wang:2024SB}. 
Despite its success, nuclear DFT faces significant challenges due to discrepancies in predictions made by different energy density functionals (EDFs). These discrepancies contribute to considerable uncertainties, especially for neutron-rich nuclei with sparse experimental data~\cite{Erler:2012} and for the equation of state of nuclear matter at densities far from the saturation point~\cite{Li:2008PR}.   As facilities for radioactive ion beams and precision measurements continue to advance~\cite{Arrowsmith-Kron:2024}, there is an increasing need to rigorously quantify the theoretical uncertainties of nuclear DFT. This effort is crucial not only for making meaningful comparisons with other models and available data,  but also for conducting systematic studies of nuclear low-lying states approaching the driplines~\cite{Rodriguez:2007PRL} and determining NMEs relevant to rare processes~\cite{Engel:2017,Belley:2024PRL}.

 In the past decade, significant progress has been made in quantifying the uncertainties of DFT predictions for nuclear ground-state bulk properties~\cite{Dobaczewski:2014,McDonnell:2015PRL,Agbemava:2019PRC}, and in identifying possible correlations between nuclear-matter quantities~\cite{Giuliani:2022} and neutron-star observables~\cite{Salinas:2023PRC}. Extending DFT to nuclear spectroscopic properties is more challenging as it usually requires going beyond the mean-field approximation~\cite{Yao:2022PPNP}. This is typically achieved through the implementation of quantum-number projections and generator coordinate method (PGCM). This  framework is known as multireference (MR) DFT~\cite{Bender:2003RMP,Niksic:2011PPNP,Robledo:2018JPG,Sheikh:2021qv,Yao:2022PPNP}, which has been successfully applied to study nuclear low-lying states~\cite{Bender:2008,Rodriguez:2010PRC,Yao:2010,Niksic:2011PPNP,Bally:2014PRL,Sheikh:2021qv} and  NMEs of $0\nu\beta\beta$ decay~\cite{Rodriguez:2010PRL,Song:2014,Yao:2022PPNP}. To date, the uncertainties in MR-DFT calculations have not been assessed due to the prohibitive computational cost of performing numerous repeated calculations with varying EDF parameters. This challenge also impedes the optimization of EDF parameters using nuclear spectroscopic data.

In this work, we develop a subspace-projected covariant density functional theory (SP-CDFT), which merges multi-reference (MR)-CDFT~\cite{Yao:2014} with the eigenvector continuation (EC) method~\cite{Frame:2018PRL} for nuclear low-lying states. The EC method, a specialized version of reduced basis methods~\cite{Quarteroni:2013,Bonilla:2022PRC}, has recently been frequently employed to emulate high-fidelity calculations in nuclear physics~\cite{Yoshida:2018PRC,Ekstrom:2019PRL,Jiang:2024}. However, it is challenging to emulate nuclear excited states~\cite{Yoshida:2022PTEP,Sun:2024}, especially in heavy nuclei where excitation energies are several orders of magnitude smaller than ground-state binding energy.  We demonstrate that our SP-CDFT can effectively reproduce MR-CDFT results for nuclear low-lying states with greatly reduced
computational effort. This approach allows us to explore the correlations among the NMEs of $0\nu\beta\beta$ decay, nuclear-matter properties, and nuclear low-lying states starting from a universal nuclear EDF, linking nuclear weak-decay properties to those of infinite nuclear matter and finite nuclei.  Our findings indicate that nuclear-matter properties at saturation density are weakly correlated with nuclear spectroscopy and $0\nu\beta\beta$ decay, whereas the latter two exhibit much stronger correlations, albeit varying with the specific nucleus. Using nuclear matter and spectroscopic data, we quantify the statistical uncertainties  of the NMEs, which are found to be much smaller than the systematic discrepancies among different nuclear models.  

 \paragraph{Emulating MR-CDFT with SP-CDFT.} We start from a relativistic EDF composed of the standard kinetic energy $\tau(\boldsymbol{r})$, electromagnetic energy  $\mathcal{E}^{\text {em}}(\boldsymbol{r})$, as well as the nucleon-nucleon ($NN$) interaction energy~\cite{Burvenich:2002PRC},
 \begin{equation}
 \label{eq:EDF}
    E[\tau, \rho,\nabla\rho; \mathbf{C}] 
    =\int d^3r \Big[\tau(\boldsymbol{r})
    +\mathcal{E}^{\text {em}}(\boldsymbol{r})
    + \sum^9_{\ell=1} c_\ell  \mathcal{E}^{NN}_\ell(\boldsymbol{r}) \Big],
\end{equation}
where $\rho$ represents different types of densities and currents. The $NN$ interaction energy is parameterized using different powers of $\rho$ with nine contact terms. The low-energy constants $c_\ell$s are collectively denoted as $\mathbf{C}=\{\alpha_S, \beta_S, \gamma_S, \delta_S, \alpha_V, \gamma_V, \delta_V, \alpha_{TV}, \delta_{TV}\}$. The subscripts $(S, V)$ indicate the scalar and vector types of $NN$ interaction vertices in Minkowski space, respectively, and $T$ for the vector in isospin space. For open-shell nuclei, the EDF (\ref{eq:EDF}) has an additional term from pairing correlation between nucleons. For the sake of simplicity, the strength parameters of pairing energy terms are fixed throughout this work. These model parameters are usually optimized to the ground-state bulk properties of finite nuclei, and infinite nuclear-matter properties around saturation density at the mean-field level~\cite{Burvenich:2002PRC,Zhao:2010PRC}.

The wave function of nuclear low-lying states in the MR-CDFT is constructed as a linear combination of mean-field configurations with projection onto good quantum numbers,
\begin{equation}
  \ket{\Psi^{JNZ}_\nu(\mathbf{C})}=\sum^{N_\mathbf{q}}_{\mathbf{q}} f^{JNZ}_{\nu}(\mathbf{q}, \mathbf{C}) \ket{JNZ; \mathbf{q}, \mathbf{C}},
\end{equation}
where the basis of many-body wave function is given by
\begin{equation}
  \ket{JNZ; \mathbf{q}, \mathbf{C}}
  =\hat P^J_{MK}\hat P^N\hat P^Z \ket{\Phi(\mathbf{q},\mathbf{C})}.
\end{equation}
Here, $\hat{P}^{J}_{MK}$ and $\hat{P}^{N, Z}$ are projection operators that select the component with angular momentum $J$ and its $z$-projection $K$, neutron number $N$, and proton number $Z$, respectively~\cite{Ring:1980}. The mean-field wave functions $\ket{\Phi(\mathbf{q},\mathbf{C})}$ are generated from a self-consistent CDFT calculation with a constraint on the mass quadrupole moment based on the EDF \eqref{eq:EDF} for a given set of parameters $\mathbf{C}$. Following our previous studies~\cite{Song:2014,Yao:2015}, we only include axially deformed and parity-conserving mean-field configurations, which are reasonable for the even-even nuclei of all concerned. In this case, we have $K=0$ for all configurations and thus omit this symbol. Pairing correlation in each configuration state is treated using the Bardeen–Cooper–Schrieffer (BCS) theory. The number of basis functions is denoted by $N_\mathbf{q}$. The mixing weight  $f^{JNZ}_{\nu}(\mathbf{q}, \mathbf{C})$ is determined with the variational principle, leading to the Hill-Wheeler-Griffin equation~\cite{Hill:1953,Griffin:1957}. For each parameter set $\mathbf{C}$, there are about $N^2_\mathbf{q}$ PGCM kernels,  
\bsub
\label{eq:GCM_kernel}
\beqn 
      {\cal N}^{\mathbf{C}}(\mathbf{q}, \mathbf{q}')
     &=& \bra{JNZ; \mathbf{q}, \mathbf{C}} JNZ; \mathbf{q}',\mathbf{C}\rangle,\\
     {\cal H}^{\mathbf{C}}(\mathbf{q}, \mathbf{q}')
    &=& \bra{JNZ; \mathbf{q}, \mathbf{C}} \hat H(\mathbf{C})\ket{JNZ; \mathbf{q}',\mathbf{C}},
\eeqn
\esub 
to be determined numerically. The Hamiltonian kernels ${\cal H}^{\mathbf{C}}(\mathbf{q}, \mathbf{q}')$ are evaluated with the generalized Wick's theorem~\cite{Balian:1969}. In particular, the energy overlap is determined with the mixed-density prescription~\cite{Bonche:1990NPA,Valor:1999NPA}. The MR-DFT framework usually meets the problems of spurious divergences~\cite{Anguiano:2001NPA,Dobaczewski:2007PRC}  and finite steps~\cite{Bender:2009PRC,Duguet:2009PRC}.  Although this issue has yet to be fully resolved~\cite{Sheikh:2021}, its impact is monitored by checking the convergent behaviors of the energies of the projected mean-field states that are ultimately used in the configuration mixing calculation for the nuclei of interest. We find that its impact on the low-lying states in MR-CDFT calculations is minor, consistent with the findings in light nuclei~\cite{Yao:2010,Zhou:2024PRC}.

To emulate the high-fidelity MR-CDFT calculation for nuclear low-lying states,  we propose a SP-CDFT($N_t, k_{\rm max}$) based on the EC method~\cite{Frame:2018PRL}. In this method, the wave function $\ket{\Psi^{JNZ}_k(\mathbf{C}_\odot)}$ of the $k$-th state for a target EDF $E[\rho, \nabla\rho; \mathbf{C}_\odot]$ is expanded in terms of the wave functions $\ket{\Psi^{JNZ}_\nu(\mathbf{C}_t)}$ (called EC basis) of the first $k_{\rm max}$ states by the $N_t$ sampling EDFs, 
\begin{equation}
\label{eq:EC_GCM_wfs}
   \ket{\bar\Psi^{JNZ}_k(\mathbf{C}_\odot)}
   =\sum^{k_{\rm max}}_{\nu=1}\sum^{N_t}_{t=1} \bar f^{JNZ}_{k, \mathbf{C}_\odot}(\nu, \mathbf{C}_t)\ket{\Psi^{JNZ}_\nu(\mathbf{C}_t)},
\end{equation}
where $k\in [1,2,\cdots, k_{\rm max}]$. The dimension of the EC basis is thus $N_{\rm EC}=N_tk_{\rm max}$.  The expansion coefficient  $\bar f^{JNZ}_{k, \mathbf{C}_\odot}(\nu, \mathbf{C}_t)$ is determined by the following generalized eigenvalue equation,
\beq 
 \sum^{k_{\rm max}}_{\nu'=1}\sum^{N_t}_{t'=1} \Bigg[ 
 \mathscr{H}^{\nu\nu'}_{tt'}(\mathbf{C}_\odot) 
 - \bar E_k^{\mathbf{C}_\odot}   \mathscr{N}^{\nu, \nu'}_{tt'} \Bigg]  \bar f^{JNZ}_{k, \mathbf{C}_\odot}(\nu', \mathbf{C}_{t'})=0,
\eeq 
where the norm and Hamiltonian kernels of EC for a target EDF $E[\rho, \nabla\rho; \mathbf{C}_\odot]$ are defined  as~\cite{Zhang:2024_Article}
\bsub
\label{eq:EC_kernel}
\beqn 
      \mathscr{N}^{\nu\nu'}_{tt'} 
     &=& \bra{\Psi^{JNZ}_\nu(\mathbf{C}_t)}\Psi^{JNZ}_{\nu'}(\mathbf{C}_{t'}\rangle,\\
   \mathscr{H}^{\nu\nu'}_{tt'}(\mathbf{C}_\odot) 
    &=& \bra{\Psi^{JNZ}_{\nu}(\mathbf{C}_t)} \hat H(\mathbf{C}_\odot) \ket{\Psi^{JNZ}_{\nu'}(\mathbf{C}_{t'})}.
\eeqn
\esub
The efficiency of the SP-CDFT is demonstrated in scenarios requiring numerous repeated MR-CDFT calculations. The speed-up factor increases almost linearly up to $10^4$ when the number of sampling EDFs reaches $10^6$, a typical value used in the assessment of statistical uncertainty~\cite{Ekstrom:2019PRL}. In other words, the SP-CDFT enables us to predict nuclear low-lying states for millions of EDFs within half an hour using a PC, a task that would otherwise take years with the high-fidelity MR-CDFT. See Ref.~\cite{Zhang:2024_Article} for more details.

\begin{figure}  
    \centering 
    \hspace{-0.1cm}\includegraphics[width=4.7cm,height=3.3cm]{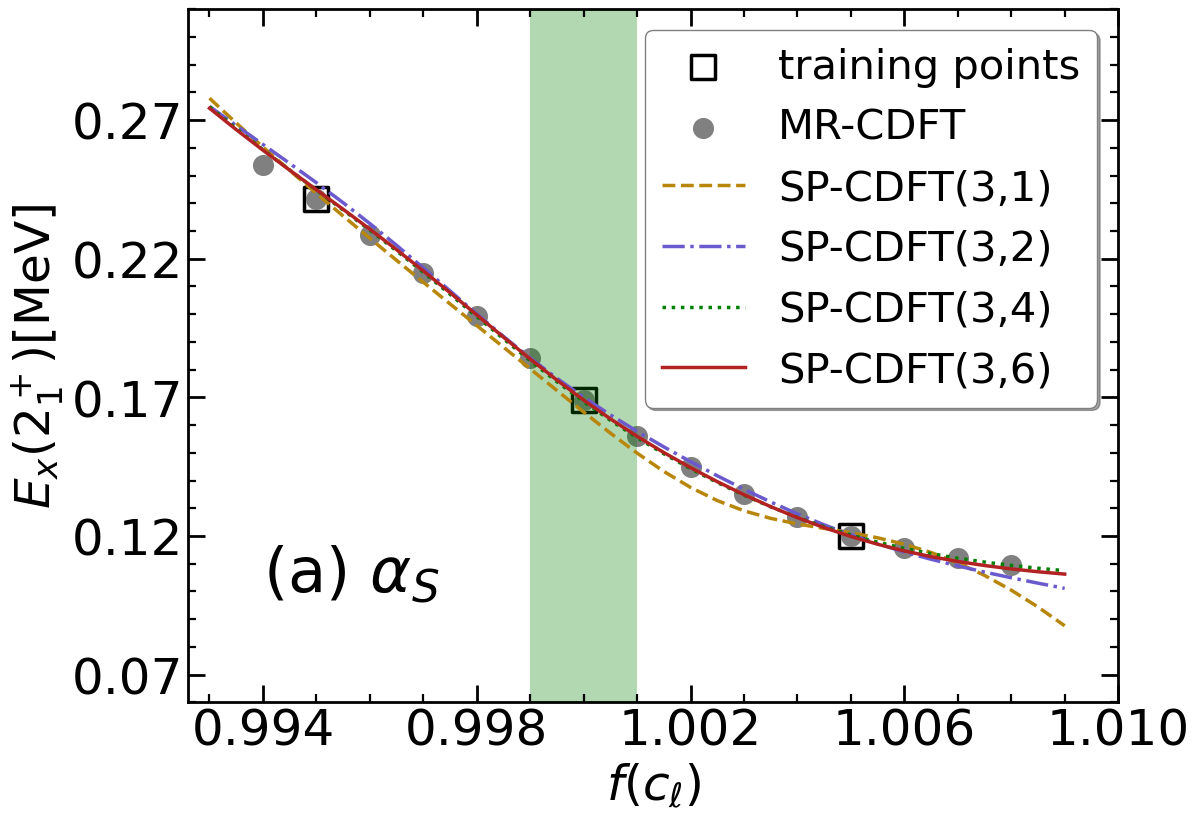} 
    \includegraphics[width=3.86cm,height=3.3cm]{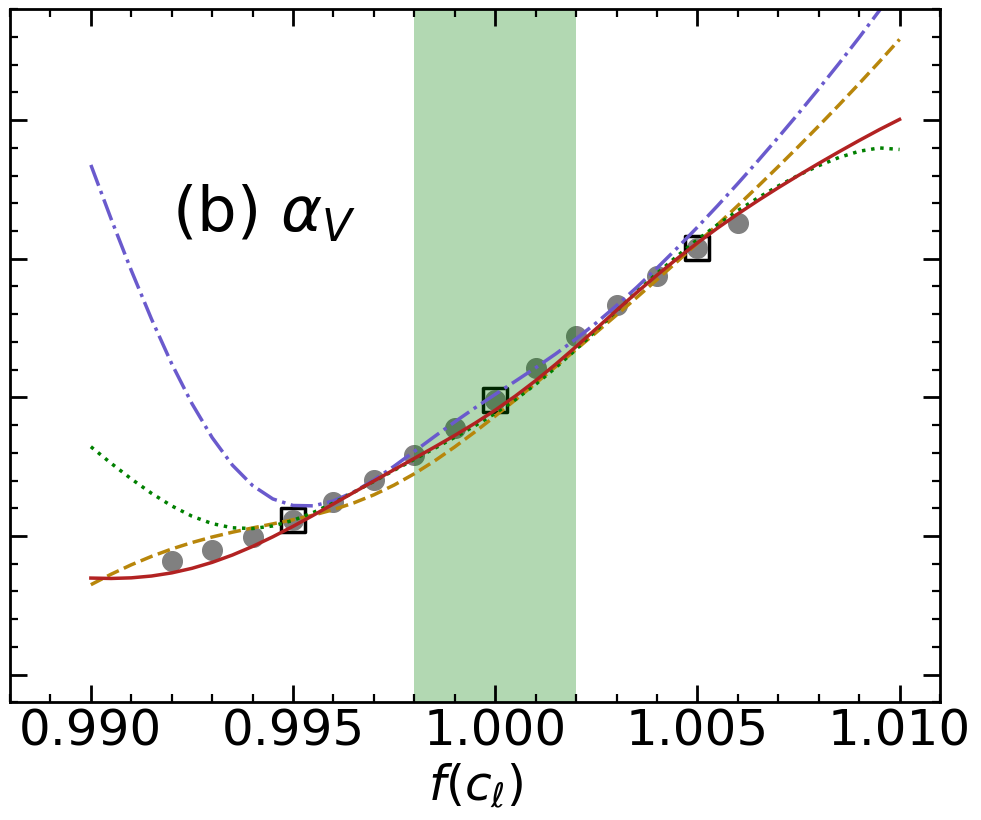} 
    \includegraphics[width=4.56cm,height=3.3cm]{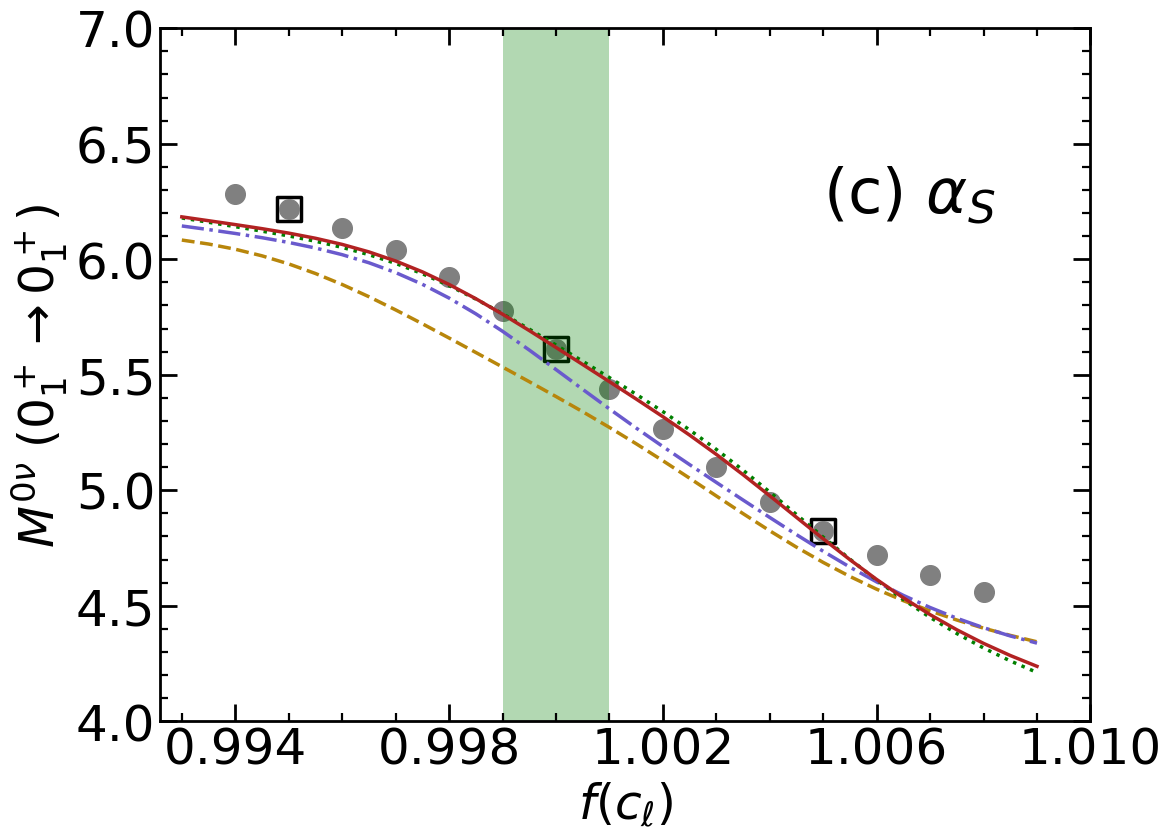} 
    \includegraphics[width=3.8cm,height=3.24cm]{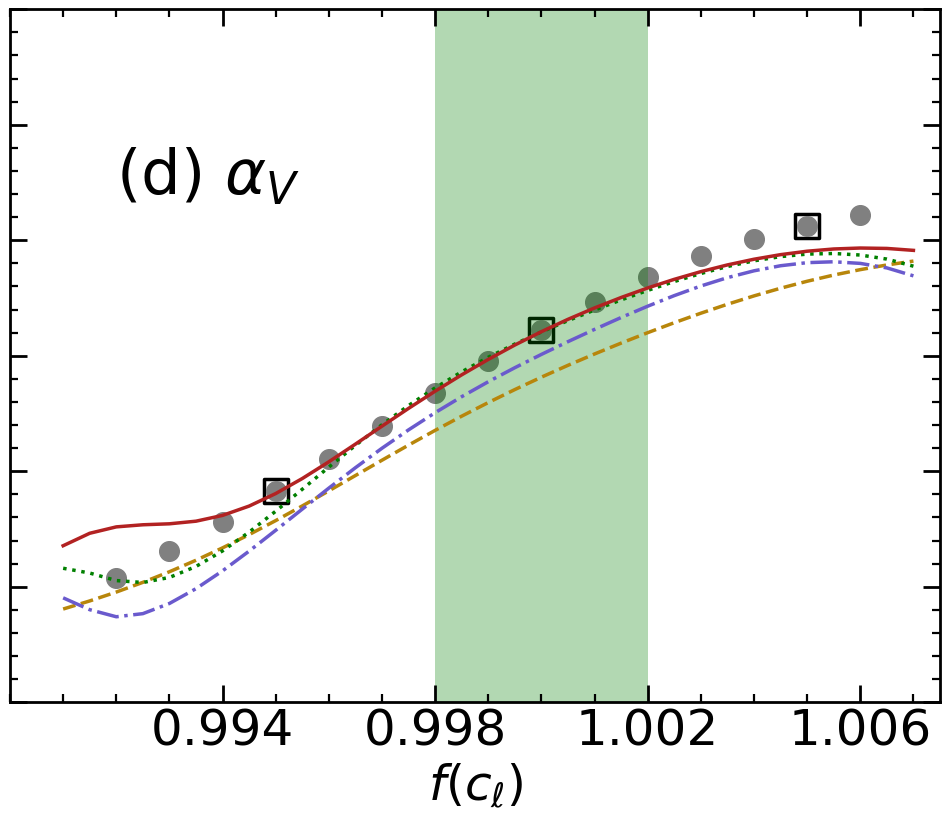} 
      \caption{(Color online) The comparison of (a, b) the excitation energies $E_x(2_1^+)$ of the first $2^+$ state in $^{150}$Nd, as well as (c, d) the NMEs of $0\nu\beta\beta$ decay, $\nuclide[150]{Nd}(0^+_1) \to \nuclide[150]{Sm}(0^+_1)$,  from both MR-CDFT and SP-CDFT($3, k_{\rm max}$) calculations with different values of $k_{\rm max}$, as a function of the scaling factors $f(c_\ell)$ of the scalar and vector coupling constants $\alpha_S$ and $\alpha_V$. The green shaded area indicates the ranges of these values in the samples.}
    \label{fig:excitation_energy_J2_Nd150} 
\end{figure}

Figure~\ref{fig:excitation_energy_J2_Nd150}(a) and (b) show the excitation energy $E_x(2_1^+)$ of the first $2^+$ state in \nuclide[150]{Nd}, obtained from both the MR-CDFT and SP-CDFT ($N_t$, $k_{\rm max}$) calculations. All results are based on the covariant energy density functional \eqref{eq:EDF} with varying values of $\alpha_S$ and $\alpha_V$. To facilitate the sampling of parameter sets, we introduce a scaling factor $f(c_\ell) = c_\ell / c^0_\ell$ for each parameter in $\mathbf{C}$, where $c^0_\ell$ is the value of the coupling constant $c_\ell$ in the PC-PK1 set~\cite{Zhao:2010PRC}, which has shown remarkable success in describing the nuclear masses of a total of 4829 isotopes~\cite{DRHBcMassTable:2024}. Since nuclear properties are mainly sensitive to these two coupling constants~\cite{Zhang:2024_Article}, we vary their values within 0.2\% during the sampling procedure. We find that SP-CDFT($N_t$, $k_{\rm max} \ge 3$) reasonably reproduces the results of  MR-CDFT calculations and generally performs better than SP-CDFT($N_t$, 1).  This finding confirms the advantage of using the extended EC scheme in Ref.~\cite{Luo:2024PRC}, which helps reduce the number $N_t$ of required training samples.  Of particular interest is the observation that the excitation energy in \nuclide[150]{Nd} increases smoothly with $\alpha_V$ and decreases with $\alpha_S$. Notably, these trends vary across different nuclei~\cite{Zhang:2024_Article}. We also checked the global performance of SP-CDFT for other spectroscopic quantities, including ground-state energies and proton radii, excitation energies of $2^+_1$ states, and $B(E2: 0^+_1 \to 2^+_1)$ for \nuclide[150]{Nd}, \nuclide[136]{Xe}, and their daughter nuclei in $0\nu\beta\beta$ decay. To this end, we sampled 14 training parameter sets and 64 testing sets in the vicinity of the PC-PK1 set using the Latin hypercube sampling method~\cite{Dutta:2020LHS}. We checked the results by sampling the parameters in the vicinity of the PC-F1~\cite{Burvenich:2002PRC}, and obtained similar conclusions. Therefore, we mainly show the results by the samples around the PC-PK1 set in this paper.   For the 64 testing sets, the relative emulator errors  are generally less than 6\%~\cite{Zhang:2024_Article}.

\begin{figure}[]
    \centering  
    \includegraphics[width=4.5cm,height=8cm]{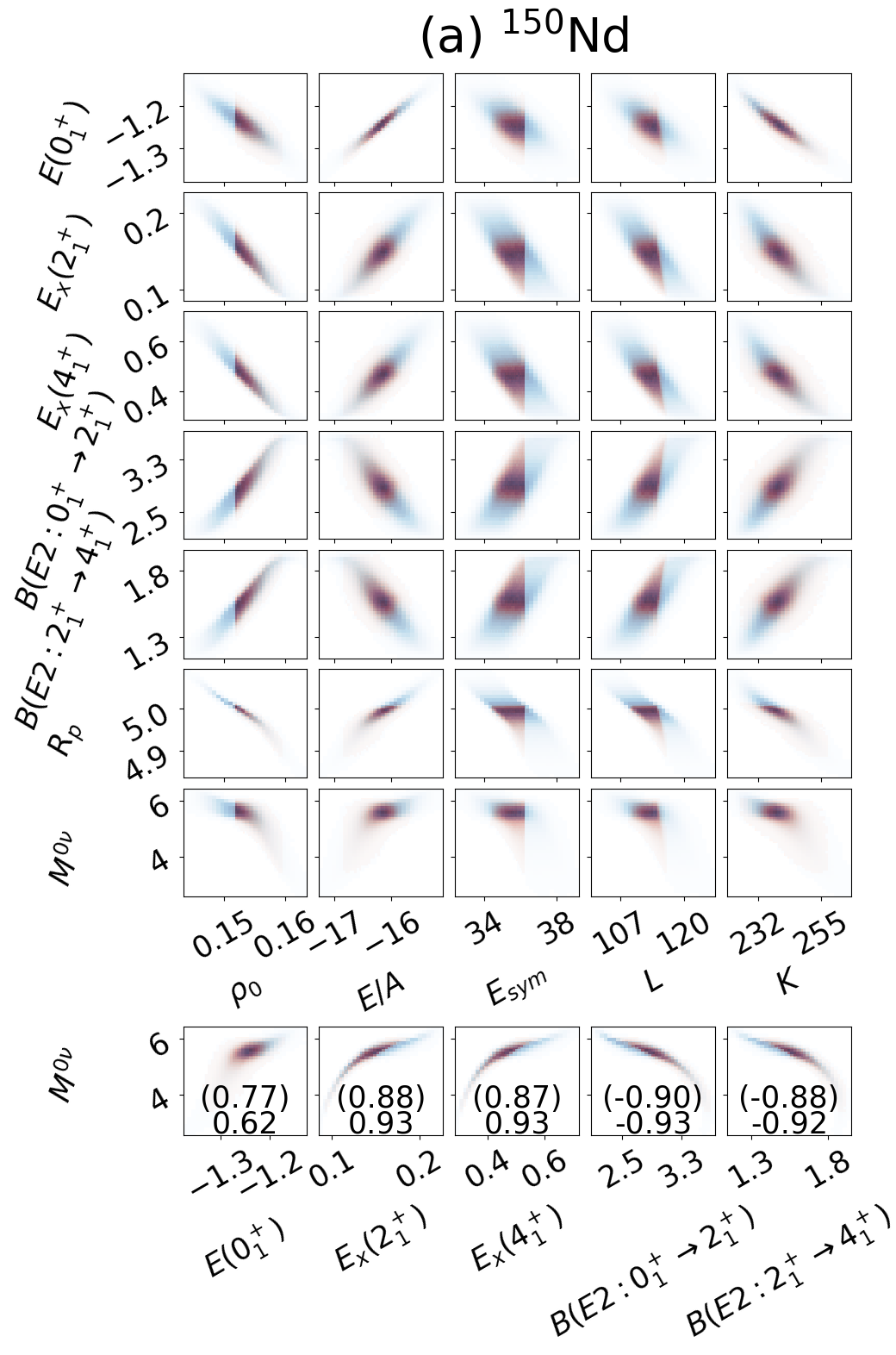} 
    \includegraphics[width=4.0cm,height=8cm]{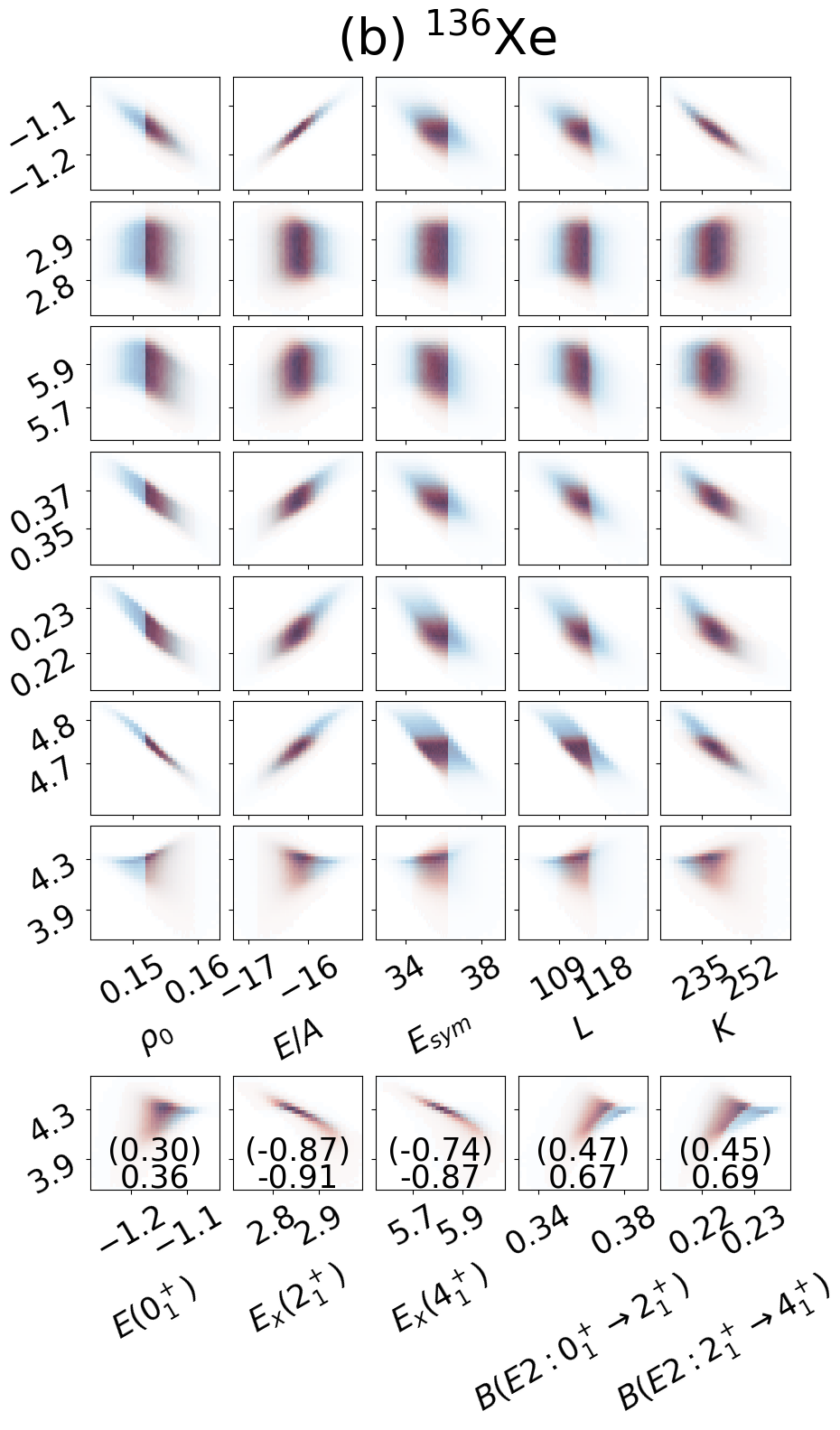} 
     \caption{(Color online) The correlation relation among different quantities, including the quantities  $\Theta_{\rm sat}=\{\rho_0, E/A, E_{\rm sym}, L, K\}$ of infinite nuclear matter at the saturation density $\rho_0$, the low-lying states and NME $M^{0\nu}$ of (a) \nuclide[150]{Nd} and (b) \nuclide[136]{Xe} from the SP-CDFT($14, 3$) calculation.  The red distributions are refined by the empirical values of  $\Theta_{\rm sat}$~\cite{Zhang:2024_Article}. The ground-state energies are in GeV, excitation energies in MeV, and $B(E2)$ in $e^2b^2$. In the bottom row, the Pearson coefficients are derived from the red distributions, while those in parentheses are based on the blue distributions. See the main text for details.}
   \label{fig:correlation}
\end{figure}

Moreover, the SP-CDFT allows us to perform a global sensitivity analysis of the low-lying states and the NME $M^{0\nu}$ to the EDF parameters. Figure~\ref{fig:excitation_energy_J2_Nd150} (c) and (d) present the comparison of the two calculations for the NME of the ground-state-to-ground-state ($0_1^+ \to 0_1^+$) $0\nu\beta\beta$ decay.
The $M^{0\nu}$ is computed using the transition operators based on the standard mechanism, see Refs.~\cite{Song:2014,Yao:2015} for details. It is important to note that the contributions from the contact transition operator~\cite{Cirigliano:2018} and two-body currents are not included in this analysis. While their omission may introduce some errors, these contributions are expected to be sub-leading for the following  reasons. Recent studies have demonstrated that the renormalizability of the transition amplitude is automatically ensured at leading order within the relativistic framework~\cite{Yang:2024PLB}, classifying the contact transition operator as a next-to-next-to-leading order (N2LO) effect in this context. Additionally, two-body currents, which appear at next-to-next-to-next-to-leading order (N3LO), are known to reduce the matrix elements by approximately 10\%~\cite{Wang:2018PRC}. It is shown in Figs.~\ref{fig:excitation_energy_J2_Nd150} (c) and (d) that the SP-CDFT(3, $k_{\rm max} \ge 3$)  can also reproduce the NME obtained from MR-CDFT calculations with reasonable accuracy.

Subsequently, we carry out a global sensitivity analysis for the parameters, in which  all parameters in $\mathbf{C}$ are independent of each other and uniformly distributed. Considering the fact that the precision of interpolation is generally higher than that of extrapolation, the range of the training points is chosen to be slightly broader than that of the sampling points for all the parameters, as explained in detail in Ref.~\cite{Zhang:2024_Article}. We find that approximately 90\% of the variance in the quantities of interest can be attributed to the parameters $\alpha_S$ and $\alpha_V$. Secondary contributions come from $\beta_s$ and $\gamma_s$, which account for most of the remaining variance. Interestingly, the sensitivity indices are consistent across different nuclei~\cite{Zhang:2024_Article}.

\paragraph{Application of the SP-CDFT to statistical analysis.}
We sampled totally about $1.3\times 10^6$ parameter sets of the EDF  by varying the nine parameters $\mathbf{C}$ around their optimal values  using quasi Monte-Carlo  sampling with a uniform distribution. To explore correlations among different quantities, we compute the properties  $\Theta_{\rm sat}=\{\rho_0, E/A, E_{\rm sym}, L, K\}$ of infinite nuclear matter at the saturation density $\rho_0$, the low-lying states, and NMEs of two heaviest candidate nuclei for $0\nu\beta\beta$ decay, where the symmetry energy  and its slope are defined as $E_\mathrm{sym} \equiv \frac{1}{2}\left.\frac{\partial^2 (E/A)}{\partial \eta^2}\right|_{\eta=0}$ and $L \equiv 3\rho
        \frac{\partial E_\mathrm{sym}}{\partial \rho}$, respectively, with $\eta = (\rho^{(n)}-\rho^{(p)})/\rho$. The total nucleon-number density is a summation of those for neutrons and protons $\rho=\rho^{(n)}+\rho^{(p)}$. 
 Figure~\ref{fig:correlation} shows the correlation relations among these quantities. In addition, we select out 457,380 samples that are able to reproduce the empirical data of the nuclear matter  simultaneously~\cite{Zhang:2024_Article}, and display  these results in Fig.~\ref{fig:correlation} as well.
 The correlations of nuclear spectroscopic quantities with $E_{\rm sym}$ and $L$ are generally weak in both cases, especially for the near-spherical nucleus \nuclide[136]{Xe}. Specifically, the excitation energy $E_x(2^+_1)$ of \nuclide[150]{Nd} is somewhat correlated to the saturation density $\rho_0$, average nucleon energy $E/A$ and incompressibility $K$. However, these correlations disappear in \nuclide[136]{Xe}.  Moreover, one observes a strong anti-correlation between the proton radius $R_p$ and  $\rho_0$,  and a linear correlation between the ground-state energy $E(0^+_1)$ and  $E/A$, consistent with the finding in a previous study based on Skyrme EDFs~\cite{Reinhard2016}. This finding justifies the use of $E/A$ and $\rho_0$ of nuclear matter to infer the prediction of EDFs for the binding energies and radii of finite nuclei, respectively~\cite{Burvenich:2002PRC,Zhao:2010PRC}.  
 
 In particular, we observe that the NME $M^{0\nu}$ of $0\nu\beta\beta$ decay is generally strongly correlated with the $E_x(2^+_1)$ and $B(E2: 0^+_1\to 2^+_1)$, even though the correlation relations are different for the prolate deformed nucleus \nuclide[150]{Nd} and spherical nucleus \nuclide[136]{Xe}.  With the refinement from the empirical values of nuclear matter at saturation density, the correlations are slightly increased, and the Pearson coefficients for $^{150}$Nd are $0.93$ and $-0.93$, respectively. For $^{136}$Xe, the Pearson coefficients become $-0.91$ and $0.67$, respectively. These numbers are much larger than the values $0.65$ and $-0.18$, respectively, found in the interacting shell model (ISM)~\cite{Horoi:2023PRC_Xe136}. In the ISM for \nuclide[48]{Ca}, these numbers are $0.23$ and $0.43$, respectively~\cite{Horoi:2022PRC_Ca48}.  It is worth noting that the Pearson coefficients between the $M^{0\nu}$ and $E_x(2^+_1)$ in  \nuclide[76]{Ge} and \nuclide[76]{Se} from the {\em ab initio} valence-space in-medium similarity renormalization group (VS-IMSRG) are found to be $0.69$ and $0.79$, respectively~\cite{Belley:2022_proceeding}.  In short, there are strong correlations between the NME $M^{0\nu}$ and the properties of nuclear low-lying states, but these correlations may vary with nuclei. The observed correlations provide a basis for employing the Bayesian model averaging method for $0\nu\beta\beta$ studies~\cite{Cirigliano:2022JPG}.

\begin{figure}[]
    \centering
    \includegraphics[width=0.8\columnwidth]{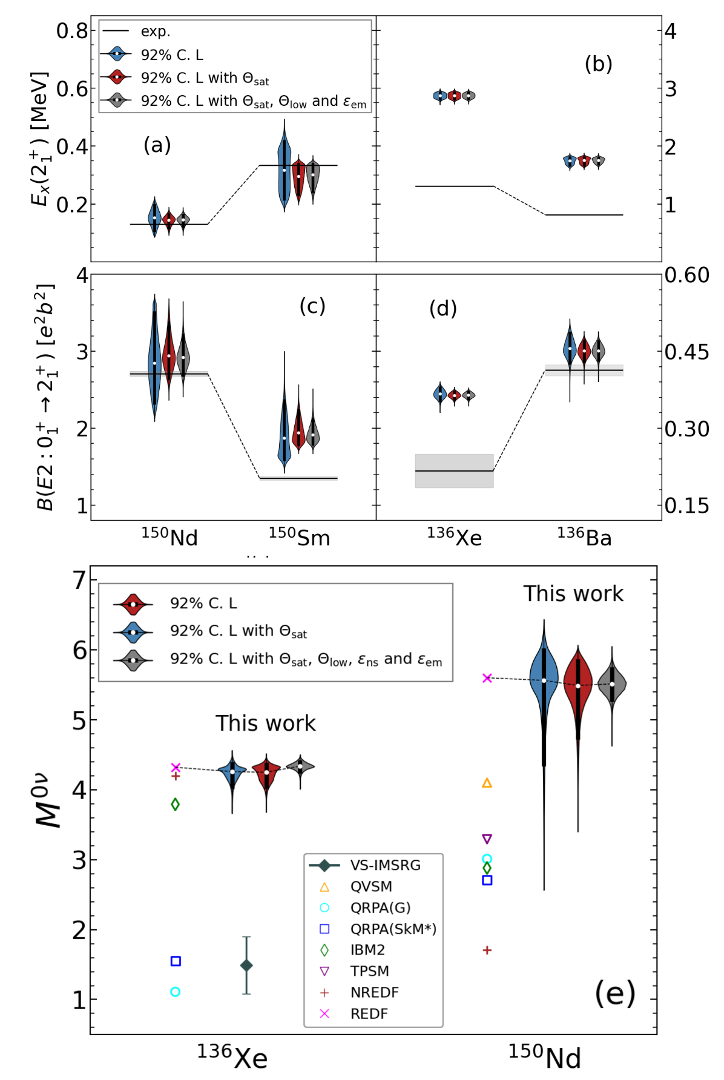} 
    \caption{(Color online) Probability distributions for the excitation energy of the $2^+_1$ state, $B(E2: 0^+_1 \to 2^+_1)$, and the NME $M^{0\nu}$  of $0\nu\beta\beta$ decay in \nuclide[150]{Nd}, \nuclide[150]{Sm}, shown without (blue) and with (red) refinement of $\Theta{\rm sat}$, as well as the posterior distribution (grey) considering $\Theta_{\rm low}$ and emulator errors ($\epsilon_{\rm em}$). The median values and uncertainties (92\% C.L.) of these distributions are displayed alongside the experimental data for comparison.  The results of $M^{0\nu}$  are compared to the values by the MR-CDFT based on the relativistic EDF PC-PK1~\cite{Yao:2015} (REDF) and those of other model calculations, including the interacting-boson model (IBM2)~\cite{Barea:2015}, MR-DFT based on a non-relativistic Gogny force (NREDF)~\cite{Rodriguez:2010PRL}, quasiparticle random-phase approximation (QRPA) based on an effective interaction of G matrix~\cite{Fang:2021PRC} and a Skyrme EDF~\cite{Mustonen:2013}, quasiparticle vacua shell model (QVSM)~\cite{Tsunoda:2023PRC},  triaxial projected shell model (TPSM)~\cite{Wang:2021}, and ab initio calculation (VS-IMSRG)~\cite{Belley:2023}. }
    \label{fig:Ex4J24_UQ}
\end{figure}

 Figure~\ref{fig:Ex4J24_UQ} shows the probability distribution of excitation energies of $2^+_1$ states and transition strengths $B(E2: 0^+_1\to 2^+_1)$ in \nuclide[150]{Nd}, \nuclide[150]{Sm}, \nuclide[136]{Xe}, and \nuclide[136]{Ba}. In the Bayesian analysis, the priors of the parameters are taken in the Gaussian form of $\chi^2$ which reflects the deviation from the optimal value~\cite{Giuliani:2022}. Besides, we also use the results (including energy per particle, pressure, speed of sound, nuclear symmetry energy and its slope) of {\em ab initio} many-body perturbation theory calculations for pure-neutron and symmetric nuclear matter with density $\rho$<0.16 fm$^{-3}$ based on the chiral $NN$ and $3N$ potentials up to N$^3$LO~\cite{Drischler:2020PRL} to weight the 457,380 samples through Gaussian form of $\chi^2$~\cite{Zhang:2024_Article}.  With the constraints from nuclear matter and the inclusion of emulator errors of SP-CDFT, independently, we obtain the posterior probability distributions for $E_x(2_1^+)$ and $B(E2)$. It is seen from Fig.~\ref{fig:Ex4J24_UQ} that the excitation energies and $E2$ transition strengths of \nuclide[150]{Nd} and \nuclide[150]{Sm} are reasonably reproduced, while the quadrupole collectivity  of \nuclide[136]{Xe} is overestimated and this is expected to be improved by including non-collective particle-hole excitation configurations in the model space.   Moreover, we find that the statistical uncertainties (92\% C.L.) from the nine parameters are within 21\% for the excitation energies and 12\% for the $E2$ transition strengths.

 In addition to nuclear matter properties, we also incorporate data on both $E_x(2^+_1)$ and $B(E2)$ to determine the posterior distribution of $M^{0\nu}$ using the Bayesian method~\cite{Giuliani:2022}, where Pearson coefficients are applied to weight their relative contributions.  The median, 4th percentile and 96th percentile of $M^{0\nu}$, derived from the probability distribution with and without the refinement from nuclear matter, as well as from the posterior distribution, are compared with those of other nuclear model calculations in Fig.~\ref{fig:Ex4J24_UQ}(e). The results indicate that incorporating nuclear matter and nuclear low-lying state information into the Bayesian method does not significantly alter the NMEs. The median value and uncertainty (92\% C.L.) for the NMEs are $4.34^{+0.09}_{-0.10}$ for \nuclide[136]{Xe} and $5.51^{+0.24}_{-0.26}$ for \nuclide[150]{Nd}, which align closely with the values of 4.32 and 5.60 obtained using the PC-PK1 functional~\cite{Yao:2015}. We also perform a statistical analysis that accounts for the correlations among different observables, obtaining an NME of $4.33^{+0.09}_{-0.11}$ for $^{136}$Xe and $5.52^{+0.22}_{-0.26}$ for $^{150}$Nd. For the samples around the PC-F1, these values become $4.21^{+0.07}_{-0.10}$ for $^{136}$Xe and $6.07^{+0.30}_{-0.32}$ for $^{150}$Nd. These two sets of results are consistent with each other when the uncertainties are taken into account.  By using the Bayesian model averaging (BMA) method for these two popular EDFs based on their predictions on the $B(E2)$ values of candidate nuclei, we obtain the NMEs of $4.34^{+0.09}_{-0.11}$ for $^{136}$Xe and  $5.52^{+0.23}_{-0.26}$ for $^{150}$Nd. 
 These results demonstrate that the statistical uncertainties in the NMEs calculated with the relativistic EDF are much smaller than the discrepancies observed among different nuclear models. It is worth noting that the statistical uncertainty shown in Fig.~\ref{fig:Ex4J24_UQ} does not include the contribution from the pairing part~\cite{Ding:2023,Lv:2023}, which may further increase the overall uncertainty.

\paragraph{Conclusions.}
In summary, we have developed a SP-CDFT to emulate MR-CDFT for nuclear low-lying states by integrating an extended eigenvector continuation method into the PGCM.  Using SP-CDFT, we have demonstrated that the NMEs of $0\nu\beta\beta$ decay in the two heaviest candidate nuclei are correlated with the properties of nuclear low-lying states, although these correlations vary among different nuclei. We derived the posterior distribution of NMEs by utilizing empirical values and ab initio calculations of nuclear-matter properties at and below saturation density, alongside data from nuclear low-lying spectroscopy.  By using the Bayesian model averaging (BMA) method in the analysis of samples in the vicinity of two popular EDFs, we obtain the NMEs of  $4.34^{+0.09}_{-0.11}$ for $^{136}$Xe and  $5.52^{+0.23}_{-0.26}$ for $^{150}$Nd.  These statistical uncertainties are notably smaller than the discrepancies observed among different nuclear models. It is important to acknowledge, however, that several sources of uncertainty remain, with systematic uncertainties being particularly challenging to quantify within the current framework. Future work should include additional generator coordinates and explore variations in both coupling constants and energy density functional forms.  The correlations observed across different nuclear models provide a foundation for applying the BMA to determine the NMEs of $0\nu\beta\beta$ decay~\cite{Snowmass_NLDBD:2022}. This approach can also be extended to PGCM calculations based on non-relativistic EDFs and chiral Hamiltonians, offering a pathway to refine nuclear EDF parameters beyond the mean-field level using nuclear spectroscopy in the future. Moreover, our method provides a tool of choice to examine the correlations between the NMEs with the observables in high-energy collisions  of candidate nuclei in the future~\cite{STAR:2024,Jia:2024,Li:2025}. 

\paragraph{Acknowledgments}

 We thank K. Hagino, H. Hergert, W.G. Jiang, C.F. Jiao, D. Lee, G. Li, and X.L. Zhang for fruitful discussions at different stages, Y.K. Wang for carefully reading the manuscript, and C. Drischler for providing nuclear-matter properties from ab initio calculations. This work is partly supported by the National Natural Science Foundation of China (Grant Nos. 12141501, 12375119 and 12405143), the Guangdong Basic and Applied Basic Research Foundation (2023A1515010936), and the Fundamental Research Funds for the Central Universities, Sun Yat-sen University. We also acknowledge the Beijing Super Cloud Computing Center (BSCC) for providing HPC resources.

 \section*{Data availability statement}
The data that support the findings of this article are openly available [66].
  
 
%

\end{document}